\documentclass{emulateapj}

\defcitealias{gak02}{G02}
\defcitealias{sa04}{SA04}

\slugcomment{Accepted for publication in The Astrophysical Journal on 
11/30/2005}

\begin{document}

\title{Time Variability in the X-ray Nebula Powered by Pulsar 
B1509$-$58}

\author{T. DeLaney and B. M. Gaensler\altaffilmark{1}}
\affil{Harvard-Smithsonian Center for Astrophysics, 60 Garden Street, 
Cambridge, MA 02138\\ tdelaney@cfa.harvard.edu, bgaensler@cfa.harvard.edu}
%\email{tdelaney@cfa.harvard.edu, bgaensler@cfa.harvard.edu}
\altaffiltext{1}{Alfred P. Sloan Research Fellow}

\and

\author{J. Arons}
\affil{Department of Astronomy, University of California, Berkeley, CA 94720\\
arons@berkeley.edu}
%\email{arons@berkeley.edu}

\and

\author{M. J. Pivovaroff}
\affil{Lawrence Livermore National Laboratory, 7000 East Ave, Livermore, 
CA 94550\\ pivovaroff1@llnl.gov}
%\email{pivovaroff1@llnl.gov}

\begin{abstract}

We use new and archival \emph{Chandra} and \emph{ROSAT} data to study the 
time variability of the X-ray emission from the pulsar wind nebula (PWN) 
powered by PSR B1509$-$58 on timescales of one week to twelve years.  There is 
variability in the size, number, and brightness of compact knots appearing 
within 20$\arcsec$ of the pulsar, with at least one knot showing a possible 
outflow velocity of $\sim0.6c$ (assuming a distance to the source of 5.2 
kpc).  The transient nature of these knots may indicate that they are 
produced by turbulence in the flows surrounding the pulsar.  A previously 
identified prominent jet extending 12 pc to the southeast of the pulsar 
increased in brightness by 30\% over 9 years; apparent outflow of material 
along this jet is observed with a velocity of $\sim0.5c$.  However, outflow 
alone cannot account for the changes in the jet on such short timescales.  
Magnetohydrodynamic sausage or kink instabilities are feasible explanations 
for the jet variability with timescale of $\sim$ 1.3$-$2 years.  An arc 
structure, located 30$\arcsec-45\arcsec$ north of the pulsar, shows 
transverse structural variations and appears to have moved inward with a 
velocity of $\sim0.03c$ over three years.  The overall structure and 
brightness of the diffuse PWN exterior to this arc and excluding the jet has 
remained the same over the twelve year span.  The photon indices of the 
diffuse PWN and possibly the jet steepen with increasing radius, likely 
indicating synchrotron cooling at X-ray energies.  

\end{abstract}


\section{Introduction}
\label{sec:int}

The pulsar B1509$-$58 and the supernova remnant (SNR) G320.4$-$1.2 
(MSH 15$-$5\emph{2}) represent one of approximately 20 known associations 
between a pulsar and a SNR.  This pulsar is one of the most energetic known, 
with a period ($P$) of 150~ms, a period derivative ($\dot{P}$) of 
1.2$\times10^{-12}$~s~s$^{-1}$, a characteristic age 
$\tau_c \equiv P/2\dot{P}\approx1700$~yr, a spin-down luminosity 
$\dot{E} \equiv 4\pi^2I\dot{P}P^{-3} \approx 1.8\times10^{37}$~ergs~s$^{-1}$ 
(for a moment of inertia $I\equiv10^{45}$~g~cm$^{2}$), and an inferred dipole 
surface magnetic field 
$B_p\approx3.2\times10^{19}(P\dot{P})^{1/2}\approx1.5\times10^{13}$~G 
\citep{kms94,lkg05}.

SNR G320.4$-$1.2 has been well studied at radio, optical, and X-ray 
wavelengths.  The radio morphology consists of a partial shell to the 
southeast and a series of bright clumps $\sim25\arcmin$ to the northwest 
\citep{gbm99} that coincide with the optical nebula RCW~89 
\citep{rcw60,shm83}.  The X-ray morphology consists of a bright, elongated 
pulsar wind nebula (PWN) with a collimated jet extending $\sim4\arcmin$ to 
the southeast \citep[hereafter G02]{shm83,gcm95,tmc96,bb97,gak02}.  To the 
northwest are thermal clumps associated with the radio clumps and RCW~89 
\citep{shm83}.  In addition, \citetalias{gak02} identified several compact 
knots close to the pulsar, plus two semicircular arcs at a distance of 
17$\arcsec$ and 30$\arcsec$ to the north of the pulsar.  The toroidal 
morphology and collimated jet are reminiscent of structures found in 
the PWNe powered by the Crab and Vela pulsars \citep{hss95,wht00,hgh01}.  
The arcs have been proposed as being due to ion-compression in the 
particle-dominated equatorial flow from the pulsar and were interpreted by 
\citetalias{gak02} as analogs of the ``wisps'' found in the Crab Nebula.  
SNR G320.4$-$1.2 has recently been observed in very high energy $\gamma$-rays 
by HESS.  The emission is elongated along the PWN axis possibly indicating 
inverse Compton scattering of relativistic electrons \citep{aaa05}.

The structures in the Crab and Vela PWNe are known to vary in brightness and 
position over short timescales (days to months).  In the case of the Crab 
Nebula, the outward moving X-ray and optical wisps (with velocity 
$v\sim0.5c$) are thought to mark the PWN termination shock, while the 
small-scale X-ray and optical knots are thought to identify unstable, 
quasi-stationary shocks in the pulsar wind \citep{hmb02}.  Radio wisps, which 
rarely correspond to optical wisps, develop and move outward at slower 
velocities ($v\sim0.3c$) and there are even more slowly moving radio features 
($v\sim10^4$ km s$^{-1}$) farther away from the pulsar \citep{bhf04}.  Recent 
observations of the wisps and polar knots in the near infrared indicate 
brightness variations on time scales as short as 20 minutes \citep{msw05}.
In the case of the Vela PWN, the X-ray arcs also move outward and vary in 
brightness by up to 30\% \citep{pks01}.  

The variability of the Vela PWN jet 
observed in X-rays has been attributed to both kink instabilities, to account 
for the dramatic shape and brightness changes over the course of days, and 
sausage instabilities, to account for the relativistically moving ``blobs'' 
\citep[$v\sim0.5c$;][]{ptk03}.  The Crab nebula jet, on the other hand, shows 
only weak X-ray morphological variations on year-long time scales 
\citep{mbp04} and relativistic outflow identified in the optical with 
$v\sim0.4c$ \citep{hmb02}.

The PWN powered by PSR B1509$-$58 represents a unique opportunity to study 
variability in PWNe.  The physical size of the PWN is approximately 10 and 
100 times larger than the Crab and Vela PWNe, respectively.  Given the 
observed variability in the Crab and Vela PWNe and the distance to PSR 
B1509$-$58 \citep[5.2$\pm1.4$~kpc;][]{gbm99}, \citetalias{gak02} predicted 
measurable variability in the PWN of PSR B1509$-$58 on time scales of a few 
years.  To that end, we obtained new \emph{Chandra X-ray Observatory} 
observations of PSR B1509$-$58 and its PWN.  In this paper we compare our new 
images to existing \emph{ROSAT} PSPC and HRI images and \emph{Chandra} ACIS-I 
images and report on the variations observed over time scales of 1 week to 12 
years.

\section{X-ray Observations and Analysis}
\label{sec:xobs}

We observed G320.4$-$1.2 with the \emph{Chandra} ACIS-I detector on 
2003~Apr~21 for a total of 9.6~ks, on 2003~Apr~28 for a total of 10~ks, and 
on 2003~Oct~18 for a total of 19.4~ks.  The data were calibrated using 
CIAO version 3.1 and CALDB version 2.28.  After filtering for good 
time intervals, the total useable exposure times were 9469~s, 9497~s, and 
18739~s for the Apr~21, Apr~28, and Oct~18 observations, 
respectively.  Exposure corrected images were constructed for each 
epoch between 0.3 and 8~keV as outlined in \citetalias{gak02}.  The telescope 
was not dithered during the 2003~Apr~28 observation so that the effects of 
the chip gaps and dead columns could not be removed at that epoch.

G320.4$-$1.2 had been previously observed with the \emph{Chandra} ACIS-I 
detector on 2000~Aug~14 for a total of 19.3~ks \citepalias{gak02}.  The 
archival data products were obtained from the \emph{Chandra} archive and were 
recalibrated using CIAO version 3.1 and CALDB version 2.28 to make use of the 
newest gain solutions and correct the geometry of the data set\footnote{For 
standard ACIS data preparation, see 
http://cxc.harvard.edu/ciao/guides/acis\_data.html.  For 
information on the geometry correction, see 
http://cxc.harvard.edu/cal/Hrma/optaxis/platescale.}.  The data were filtered 
for good time intervals resulting in a final exposure time of 17863~s and an 
exposure corrected image was made as outlined above.

We compare these \emph{Chandra} data to archival observations with the 
\emph{R\"{o}ntgen Satellite} \citep[\emph{ROSAT};][]{t82,p87}.  G320.4$-$1.2 
was observed with the \emph{ROSAT} PSPC detector on 1991~Feb~22 -- Mar~8 
and 1992~Feb~25 -- 27 for a total of 9.1~ks \citep{gcm95,tmc96}.  Data 
were also taken with the \emph{ROSAT} HRI detector on 1994~Feb~10 -- 23 
and 1994~Sep~15 for a total of 22.5~ks \citep{bb97}.  The archival data 
products were obtained from NASA's 
HEASARC\footnote{http://heasarc.gsfc.nasa.gov/}.  Exposure corrected images 
were made using the \emph{Xselect} utility and the \emph{hriexpmap} and 
\emph{pcexpmap} tasks within version 5.3.1 of the FTOOLS package.  The 
exposure times of the PSPC and HRI data are approximately matched to 
produce the same number of counts on the two images.  To facilitate 
comparison to the \emph{Chandra} images, exposure corrected images were also 
made from 2~ks each of the 2000~Aug~14 and 2003~Oct~18 \emph{Chandra} data 
sets to match the energy range ($\sim 0.1-2.5$ keV) and number of 
counts\footnote{Version 3.6a of the PIMMS program 
(http://cxc.harvard.edu/toolkit/pimms.jsp) was used to determine the
\emph{Chandra} exposure time needed to match the number of counts on the 
\emph{ROSAT} image for the diffuse PWN and knots in the RCW~89 region.} on 
the \emph{ROSAT} PSPC and HRI images.

The four \emph{Chandra} images were registered to each other using a 
brightness-weighted mean of PSR 
B1509$-$58, the star Muzzio 10 \citep[located at (J2000)
R.A. $15^{\mbox{h}}13^{\mbox{m}}55\fs2$, 
decl.  $-59\degr07\arcmin51\farcs6$;][]{m79}, and four other background 
sources in the 
field (located at (J2000) R.A. $15^{\mbox{h}}13^{\mbox{m}}41^{\mbox{s}}$,
decl. $-59\degr11\arcmin45\arcsec$;
R.A. $15^{\mbox{h}}14^{\mbox{m}}00^{\mbox{s}}$, 
decl. $-59\degr12\arcmin38\arcsec$; 
R.A. $15^{\mbox{h}}14^{\mbox{m}}05^{\mbox{s}}$, 
decl. $-59\degr14\arcmin40\arcsec$; 
R.A. $15^{\mbox{h}}14^{\mbox{m}}32^{\mbox{s}}$, 
decl. $-59\degr08\arcmin09\arcsec$).  The relative registration error is 
$<0\farcs1$ with rotation constrained to $<1\degr$.  Although the pulsar does 
not have a significant proper motion \citep[$\mu_{\alpha}<39$ mas yr$^{-1}$, 
$\mu_{\delta}<52$ mas yr$^{-1}$,][]{gbm99}, its suitability for 
registration purposes is in question because the strong pile-up enhances the 
azimuthal brightness asymmetry that results from a misalignment of the 
telescope mirrors in the innermost shell \citep{jdt00,pks01}.  Therefore, we 
also determined the registration solution excluding the pulsar.  Again using 
a brightness-weighted mean, we find a relative registration error $<0\farcs1$ 
for the 2000~Aug 14 and 2003~Oct~18 images and $<0\farcs25$ for the two 2003 
Apr images.

Spectra were extracted for selected sources of interest using the FTOOLS 
\emph{Xselect} utility for the \emph{ROSAT} PSPC data and the CIAO 
\emph{acisspec} script for the \emph{Chandra} data.  Background spectra were 
extracted from an annular region surrounding each source of interest.  The 
spectra were rebinned to have a minimum of 20 counts per spectral channel.  
XSPEC version 11.3.1 was used for the spectral analysis.  Unless otherwise 
noted, we present the 90\% confidence limit for errors.

All physical size and distance calculations assume that the source is 5.2 kpc 
away \citep{gbm99}. 

\section{Summed \emph{Chandra} Image}

\begin{figure*}[ht]
\epsscale{1.0}
\plotone{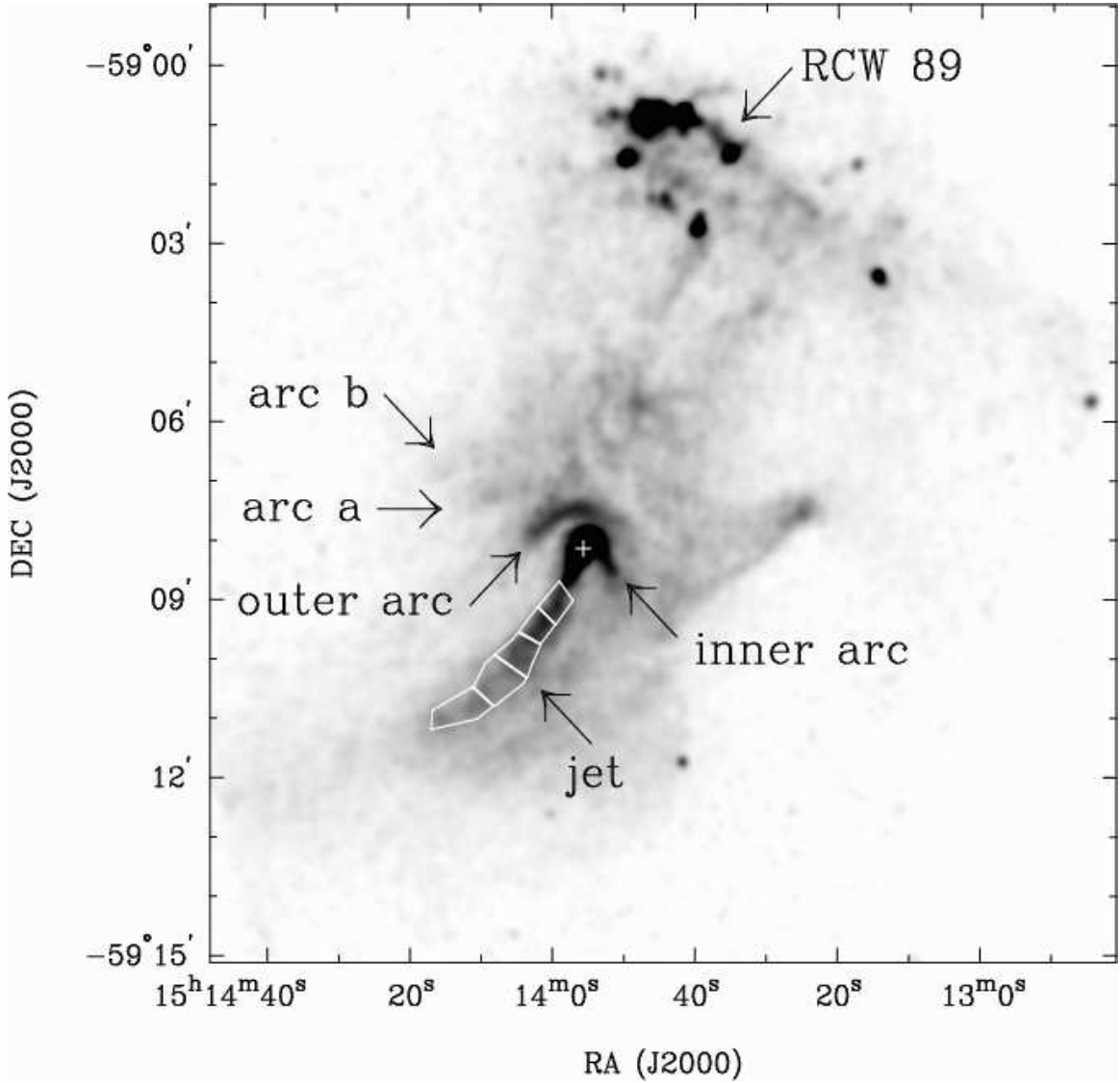}
\caption{Summed 50 ks \emph{Chandra} image of G320.4$-$1.2 over the energy 
range 0.3$-$8.0 keV.  The image has been exposure-corrected and convolved 
with a Gaussian of FWHM 10$\arcsec$.  The transfer function is linear.  On 
this and subsequent images, the pulsar position is indicated by the white 
cross.  Overlaid on the jet are the regions used to determine the radial 
variation of the photon index as indicated in Table \ref{radtable}.
\label{allim}}
\end{figure*}

Shown in Figure \ref{allim} is a summed and exposure corrected image of 50~ks 
of the \emph{Chandra} data \emph{excluding} the 2003~Apr~28 observation, 
which was not dithered.  The energy range is 0.3$-$8.0~keV and the image has 
been convolved with a Gaussian of FWHM~10$\arcsec$.  The pulsar position is 
indicated by the white cross.  In addition to the jet and inner and outer 
arcs (G02 features ``C'', ``5'', and ``E'', respectively), we identify two 
faint arc-like structures at distances of $\sim$~110$\arcsec$ and 
160$\arcsec$ from the pulsar.  These newly identified features are labelled 
as ``arc~a'' and ``arc~b'' and are $\sim$20$\arcsec$ in width.  Arc~a can be 
identified in the individual 20~ks 2000~Aug and 2003~Oct images and also 
appears in more recent \emph{Chandra} images (P. Slane, private 
communication).  The newer \emph{Chandra} data will require more analysis 
before we can determine if arc~b is also a robust structure.  

\section{Time Evolution}

\subsection{Flux of the Diffuse PWN}
\label{sec:dpwn}

To determine if the total flux of the diffuse PWN varied over time, we 
performed simultaneous absorbed power law fits to the \emph{ROSAT} PSPC and 
the \emph{Chandra} data sets, requiring the same absorbing column ($N_H$) and 
photon index ($\Gamma$) for all data sets, but allowing the normalization 
to vary between epochs (to account for both calibration uncertainties and 
actual, evolutionary brightness variations).  We define the \emph{diffuse} 
PWN to be the low brightness emission surrounding the pulsar, interior to 
the RCW 89 region to the northwest, and extending slightly beyond the jet to 
the southeast.  We specifically exclude the jet and inner and outer arc 
structures from this definition.  The exact region used for the analysis is 
the same as shown in Figure~7 of \citetalias{gak02}.  The best fit values are 
$N_H=(8.6\pm0.2)\times 10^{21}$~cm$^{-2}$ and $\Gamma=1.97\pm0.03$.  Table 
\ref{fluxes} shows the unabsorbed fluxes obtained from the simultaneous 
spectral fit for the energy range 0.5$-$10.0~keV.\footnote{Although the 
\emph{ROSAT} PSPC data do not extend above 2.5~keV, fluxes were extrapolated 
up to 10~keV using a model spectrum with the same photon index and 
normalization found from the spectral fits.}  Note that the loss of dithering 
for the 2003~Apr~28 observation resulted in lower flux values for the diffuse 
PWN and jet.  The total unabsorbed flux of the diffuse PWN has remained 
steady at $\approx5.8\times10^{-11}$~erg~cm$^{-2}$~s$^{-1}$ from 1991/1992 
to 2003.  

\begin{deluxetable*}{lcclcc}
\tabletypesize{\footnotesize}
\tablecaption{Spectral Fits to Various Subregions of the Source    
\label{fluxes}}
\tablewidth{0pt}
\tablehead{
\colhead{} &
\colhead{} &
\colhead{} &
\colhead{} &
\colhead{} &
\colhead{$F_X$\tablenotemark{c}} \\
\colhead{Region} &
\colhead{S/I\tablenotemark{a}} &
\colhead{R/C\tablenotemark{b}} &
\colhead{Epoch} &
\colhead{$\Gamma$} &
\colhead{(10$^{-12}$ erg cm$^{-2}$ s$^{-1}$)}}
\startdata
Diffuse PWN & S & R & 1991/1992   & 1.97$\pm$0.03 & 57$\pm$3 \\
            & S & C & 2000 Aug 14 &               & 58$\pm$3 \\
            & S & C & 2003 Apr 21 &               & 58$\pm$3 \\
            & S & C & 2003 Apr 28\tablenotemark{d} & & 53$\pm$3 \\
            & S & C & 2003 Oct 18 &               & 58$\pm$3 \\
            & I & C & 2000 Aug 14 & 1.97$\pm$0.03 & 58$\pm$3 \\
            & I & C & 2000 Apr 21 & 1.98$\pm$0.03 & 59$\pm$3 \\
      & I & C & 2003 Apr 28\tablenotemark{d} & 1.96$\pm$0.03 & 53$\pm$3 \\
            & I & C & 2003 Oct 18 & 1.96$\pm$0.03 & 58$\pm$3 \\
Jet         & S & R & 1991/1992   & 1.64$\pm$0.07 & 4.1$\pm$1.0 \\
            & S & C & 2000 Aug 14 &               & 5.5$\pm$0.6 \\
            & S & C & 2003 Apr 21 &               & 5.2$\pm$1.0 \\
            & S & C & 2003 Apr 28\tablenotemark{d} & & 4.1$\pm$1.0 \\
            & S & C & 2003 Oct 18 &               & 5.4$\pm$0.6 \\
            & I & C & 2000 Aug 14 & 1.62$\pm$0.11 & 5.5$\pm$0.6 \\
            & I & C & 2003 Apr 21 & 1.53$\pm$0.11 & 5.3$\pm$1.0 \\
      & I & C & 2003 Apr 28\tablenotemark{d} & 1.79$\pm$0.13 & 4.0$\pm$1.0 \\
            & I & C & 2003 Oct 18 & 1.65$\pm$0.11 & 5.4$\pm$0.6 \\
Outer Arc   & S & R & 1991/1992   & 1.64$\pm$0.09 & 4.4$\pm$1.0 \\
            & S & C & 2000 Aug 14 &               & 3.7$\pm$0.6 \\
            & S & C & 2003 Apr 21 &               & 3.4$\pm$1.0 \\
            & S & C & 2003 Apr 28 &               & 3.3$\pm$1.0 \\
            & S & C & 2003 Oct 18 &               & 3.5$\pm$0.6 \\
\enddata
\tablecomments{Uncertainties are all at 90\% confidence.  The models used are 
power law of the form $f_E\propto E^{-\Gamma}$ where $\Gamma$ is the photon 
index and the integrated flux is 
$F_X \equiv F(E1,E2) = \int_{E1}^{E2} E f(E) dE$. 
All models assume interstellar absorption using the cross sections of 
\citet{bm92}, assuming solar abundances.}
\tablenotetext{a}{``S'' indicates the results of simultaneous fits to 
multi-epoch data requiring the same $\Gamma$, holding $N_H$ fixed at 
$8.6\times10^{21}$~cm$^{-2}$ and allowing the normalization to vary between 
epochs as outlined in \S \ref{sec:dpwn}.  ``I'' 
indicates a fit to an individual epoch holding $N_H$ fixed at 
$8.6\times10^{21}$~cm$^{-2}$.}
\tablenotetext{b}{``R'' indicates \emph{ROSAT} observations, ``C'' indicates 
\emph{Chandra} observations.}
\tablenotetext{c}{Fluxes are for the energy range 0.5$-$10.0 keV, and 
have been corrected for interstellar absorption.}
\tablenotetext{d}{The telescope was not dithered during the 2003~Apr~28 
observation resulting in a decrease in measured flux for those structures 
that were intersected by chip gaps and dead columns.}
\end{deluxetable*}

The value of the absorbing column does not match that of \citetalias{gak02} 
within the errors.  We fitted just the recalibrated 2000 data set studied by 
\citetalias{gak02} and found $N_H=(8.2\pm0.3)\times10^{21}$~cm$^{-2}$ and 
$\Gamma=1.91\pm0.04$, which is consistent with the multi-epoch fit.  We also 
fixed $N_H$ at $9.5\times10^{21}$~cm$^{-2}$ and $\Gamma$ at 2.05 (as found by 
\citetalias{gak02}) to determine what, if any difference there is in total 
flux.  For the energy range 0.5$-$10.0 keV, we obtained an unabsorbed 
flux of $5.8\times10^{-11}$~erg~cm$^{-2}$~s$^{-1}$ for both sets of values 
for the absorbing column and the photon index indicating that these parameters 
are sufficiently degenerate that the slightly different values do not 
significantly affect the fitted flux.  This flux is within the error of that 
reported by \citetalias{gak02}.  We conclude that the mismatch in our 
absorbing column and photon index values compared to the \citetalias{gak02} 
results is most likely due to updated \emph{Chandra} calibration solutions.  
Unless otherwise indicated, spectral fits in this paper hold $N_H$ fixed at 
the simultaneous epoch fit value of 8.6$\times10^{21}$~cm$^{-2}$.  Fluxes are 
reported for an energy range of 0.5$-$10.0 keV and have been corrected for 
interstellar absorption.

\subsection{Flux of the Pulsar}

To determine if the pulsar flux has varied over time, we compared the 
background subtracted count rate from a $1\farcs4$ radius region around the 
pulsar for each \emph{Chandra} epoch over the energy ranges 0.5$-$10.0 keV 
and 2.0$-$10.0~keV.  Note that these count rates do not reflect the true flux 
from the pulsar because of pile-up.  We chose the two different energy ranges 
to determine if any variation in count rate could be due to pile-up 
differences between epochs.  A reduction in pile-up might result from the 
extra absorption at low energies caused by the time-dependent ACIS filter 
contamination.  The count rates are shown in Table \ref{psrcts}.  In 2000, 
the pulsar count rates between 0.5$-$10.0 keV and between 2.0$-$10.0~keV were 
$\sim$8\% less than in 2003, representing less than a 3$\sigma$ difference.  
Because the difference is approximately the same for the two energy ranges, 
it is unlikely that we have a significant bias from a reduction of pile-up at 
low energies.  We also determined the flux by performing a power law fit to 
each of the \emph{Chandra} data sets correcting for interstellar absorption 
and pile-up \citep{dav01}.  We held $N_H$ fixed at 
$8.6\times 10^{21}$~cm$^{-2}$ and we held $\Gamma$ fixed at 1.19 
\citep{cmm01} for each epoch but allowed the normalization and pile-up grade 
morphing parameter to vary.  As shown in Table \ref{psrcts}, these fits 
indicate that the flux of the pulsar has remained steady at 
$\sim5.1\times10^{-11}$ erg cm$^{-2}$ s$^{-1}$.

\begin{deluxetable*}{lccc}
\tabletypesize{\footnotesize}
\tablecaption{Detected Count Rate and Flux for PSR B1509$-$58    
\label{psrcts}}
\tablewidth{0pt}
\tablehead{
\colhead{} &
\colhead{Total Count Rate\tablenotemark{a}} &
\colhead{Hard Count Rate\tablenotemark{b}} &
\colhead{Flux\tablenotemark{c}} \\
\colhead{Epoch} &
\colhead{(counts s$^{-1}$)} &
\colhead{(counts s$^{-1}$)} &
\colhead{(10$^{-11}$ erg cm$^{-2}$ s$^{-1}$)}}
\startdata
2000 Aug 14 & 0.130$\pm$0.005 & 0.103$\pm$0.005 & 5.0$\pm$1.8 \\
2003 Apr 21 & 0.144$\pm$0.008 & 0.115$\pm$0.008 & 5.1$\pm$1.8 \\
2003 Apr 28 & 0.138$\pm$0.008 & 0.108$\pm$0.008 & 5.1$\pm$1.8 \\
2003 Oct 18 & 0.142$\pm$0.005 & 0.111$\pm$0.005 & 5.2$\pm$1.8 \\
\enddata
\tablenotetext{a}{Count rates are for the energy range 0.5$-$10.0 keV.  They 
have been background subtracted but have \emph{not} been corrected for 
pile-up.  Errors are from Poisson statistics.}
\tablenotetext{b}{Count rates are for the energy range 2.0$-$10.0 keV.  They 
have been background subtracted but have \emph{not} been corrected for 
pile-up.  Errors are from Poisson statistics.}
\tablenotetext{c}{Fluxes are for the energy range 0.5$-$10.0 keV, and 
have been corrected for interstellar absorption and pile-up.  Uncertainties 
are all at 90\% confidence.  The models used are power law of the form 
$f_E\propto E^{-\Gamma}$ where $\Gamma$ is the photon index and has been held 
fixed at 1.19 \citep{cmm01} and the integrated flux is $F_X \equiv F(E1,E2) = 
\int_{E1}^{E2} E f(E) dE$.  All models assume interstellar absorption 
using the cross sections of \citet{bm92}, assuming solar abundances with 
$N_H$ held fixed at $8.6\times10^{21}$~cm$^{-2}$.  We have implemented the 
fast pile-up algorithm of \citet{dav01}.}
\end{deluxetable*}

\subsection{Small-Scale Structure Near the Pulsar}
\label{sec:knots}

\begin{figure}[hb]
\epsscale{1.12}
\plotone{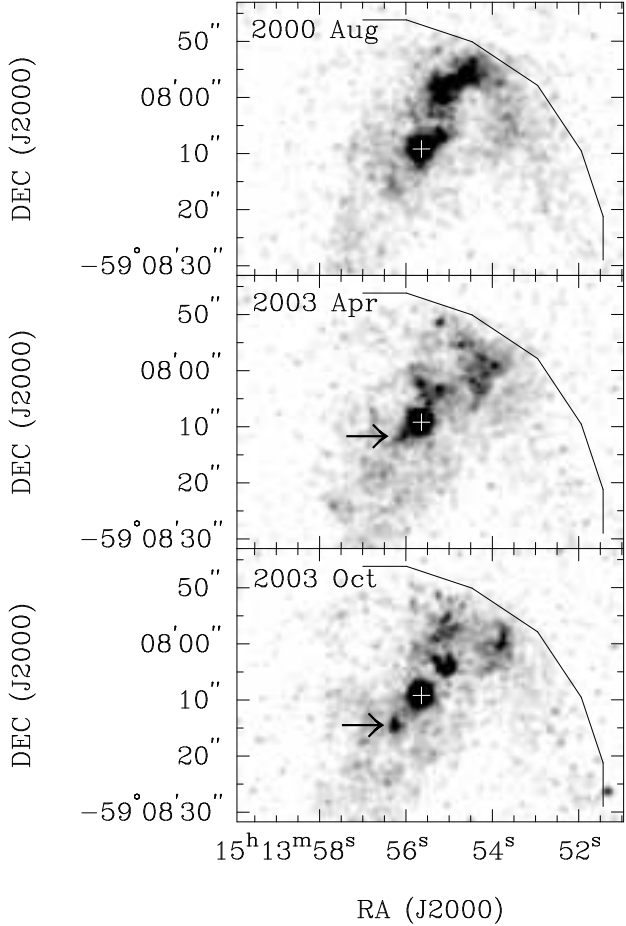}
\caption{\emph{Chandra} images showing changes in the environment immediately 
surrounding the pulsar.  The 2003~Apr image combines both April observations.  
The images have been convolved with a Gaussian of FWHM 1$\arcsec$.  The 
energy range is 0.3$-$8.0~keV.  The grayscale range is linear and ranges 
from 0 to 3\% of the peak value on all three images.  The black arc indicates 
the location of the outer edge of the inner arc (see Figure \ref{allim}).  
Arrows on 2003 images indicate a possible knot in motion.  
\label{smallscale}}
\end{figure}

Figure \ref{smallscale} shows changes in the structure immediately 
surrounding the pulsar between 2000~Aug and 2003~Oct\footnote{Note that the 
small-scale structure identified here was beyond the resolution capability of 
the \emph{ROSAT} detectors.}.  In 2000, the structure consisted of four 
compact knots $<3\arcsec$ across (corresponding to a physical size 
of 0.1$-$0.2~pc) and located along the jet axis between distances of 
$\sim3\arcsec-17\arcsec$ on both sides of the pulsar.  Three years later 
in 2003~Apr, this structure has completely changed, with approximately nine 
small, unresolved knots ($<0.5\arcsec$, $<0.01$pc) again located along the 
jet axis between distances of $\sim3\arcsec-17\arcsec$ on both sides of the 
pulsar.  No apparent changes were evident between 2003~Apr~21 and 
2003~Apr~28.  Approximately six months later in 2003~Oct, the structure has 
again changed, with six knots ranging in size from $0\farcs5$ to 4$\arcsec$ 
and again located along the jet axis within 17$\arcsec$ of the pulsar.  In all 
cases, the knot activity is greater to the northwest than to the southeast 
and in no case are knots found beyond the inner arc at a radius of 
17$\arcsec$.  

There is the possibility that some of the knot variability we observe could 
be due to artifacts.  It is known that artificial structures may appear near 
sources with strong pile-up.  However, we believe that most of the 
variability is real because the typical count rates are 
$\sim$~0.005-0.05~counts~s$^{-1}$ (corresponding to 100-1000 counts on the 
2000~Aug and 2003 
Oct images) -- higher than would be expected for artificial structures.  
Another indication that the small-scale knots are genuine comes from our 
Monte Carlo simulations discussed in \S \ref{sec:arcs}.  In each of our ten 
Poisson noise simulations, prior to smoothing to 10$\arcsec$, we recover all 
of the knot structures.  Furthermore, in analyzing their knot 1, G02 
specifically ruled out asymmetries in the wings of the point-spread 
function and pulsar photons assigned to the wrong location on the sky as 
possible explanations.  Finally, the small-scale knots are apparent on 
unsmoothed images and are not due to the 1$\arcsec$ FWHM Gaussian convolution 
applied to Figure \ref{smallscale}.  

The knot structure changes so drastically that identifying motion is 
practically impossible.  However, if the knot to the southeast of the pulsar 
in 2003~Apr and 2003~Oct is indeed the same feature (identified with arrows 
on Figure \ref{smallscale}), then the 4$\arcsec$ motion results in a velocity 
of $\sim0.6c$ (assuming outflow along the jet with a 30$\degr$ 
inclination\footnote{\citetalias{gak02} derive a 30$\degr$ jet inclination to 
the line-of-sight based on Doppler boosting and radio polarization arguments, 
while \citet{ykk05} derive an inclination angle $>50\degr$ based on an 
interaction between the unseen northwest jet and the RCW~89 region.} to the 
line of sight and correcting for relativistic Doppler boosting).

\subsection{The Jet}
\label{sec:jet}

We performed simultaneous absorbed power law fits to the \emph{ROSAT} PSPC 
and \emph{Chandra} data for the jet in the same manner as for the diffuse 
PWN in \S \ref{sec:dpwn}.  We fixed $N_H$ at $8.6\times10^{21}$~cm$^{-2}$ 
and required the same photon index for all the data sets but allowed the 
normalization to vary between epochs.  The photon indices and fluxes are 
indicated in Table \ref{fluxes}.  As noted by \citetalias{gak02}, the photon 
index of the jet is flatter than the diffuse PWN.  From the spectral fits, we 
note that the jet brightened by $\sim$30\% between the \emph{ROSAT} PSPC 
observation in 1991/1992 and the \emph{Chandra} observations, however the 
flux remained steady from 2000 to 2003.  The flux difference between the 
1991/1992 \emph{ROSAT} data and the 2000 \emph{Chandra} data is slightly 
greater than the 90\% confidence limit.  This flux difference might be caused 
by the differing spectral responses of \emph{ROSAT} and \emph{Chandra}, 
however this is presumably accounted for by the corresponding auxiliary 
response files.  We can also exclude a systematic uncertainty in the 
calibration of the two observatories as contributing to the flux difference 
by examining the flux of the Diffuse PWN.  As seen in Table \ref{fluxes}, the 
flux from the extended emission remains constant as expected.  

\begin{figure*}[ht]
\epsscale{1.0}
\plotone{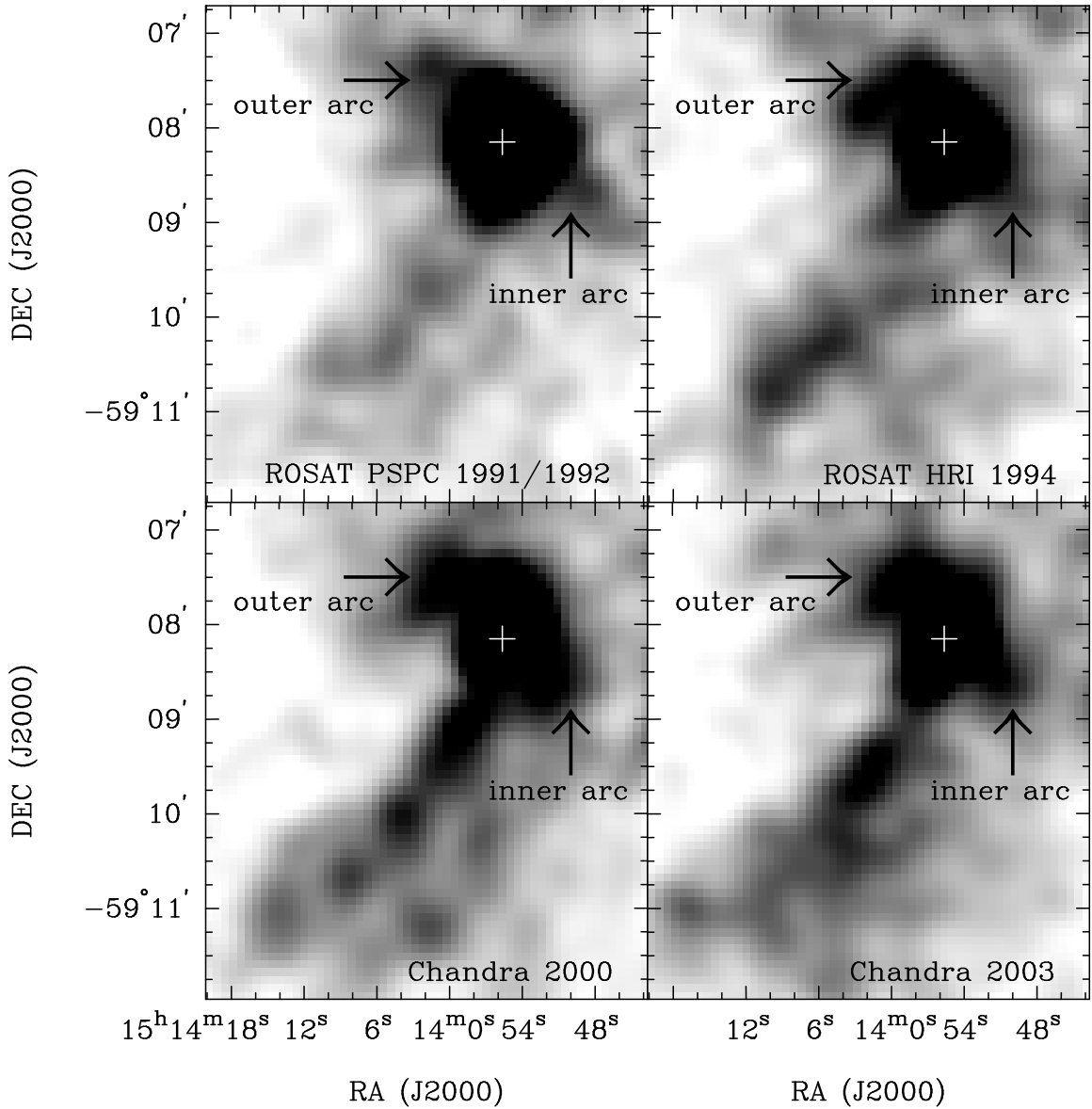}
\caption{\emph{Chandra} and \emph{ROSAT} images showing changes in the outer 
arc and jet from 1991/1992 to 2003.  To facilitate comparison between 
different observatories, the images are matched in resolution (25$\arcsec$), 
energy (0.5$-$10.0 keV), and counts as described in \S \ref{sec:xobs} and 
\S \ref{sec:jet}.  The grayscale is linear and ranges from 2.2\% to 13\% of 
the peak value for the \emph{ROSAT} PSPC and \emph{Chandra} images and 
3.9\% to 13\% of the peak value on the \emph{ROSAT} HRI image to account for 
the higher background of the HRI detector.  
\label{rcjet}}
\end{figure*}

We also compared the 1991/1992 \emph{ROSAT} PSPC image, 1994 \emph{ROSAT} HRI 
image and 2000~Aug and 2003~Oct \emph{Chandra} images of the jet as shown in 
Figure \ref{rcjet}.  To facilitate comparison between different observatories 
that might arise from the different responses of \emph{ROSAT} and 
\emph{Chandra}, the \emph{Chandra} images are produced from events restricted 
to the 0.1-2.5 keV energy band of \emph{ROSAT}.  The resultant \emph{Chandra} 
images, as well as the \emph{ROSAT} HRI image, are then convolved with a 
Gaussian of FWHM 25$\arcsec$ resolution to match the angular resolution of 
the \emph{ROSAT} PSPC.  Finally, to ensure each data set has approximately 
the same number of total counts as the \emph{ROSAT} PSPC and HRI observations, 
only a fraction of the \emph{Chandra} observations were used to generate the 
images as discussed in \S \ref{sec:xobs}.  The images clearly show 
variability in the jet structure and brightness over a twelve-year period, 
supporting the flux measurements from the spectral fits.  Indeed, the jet is 
dim and poorly defined in 1991/1992.  In 1994, however, part of the jet 
structure $\sim3\arcmin$ (corresponding to a distance of 9 pc, correcting for 
a jet inclination of 30$\degr$ to the line of sight) to the southeast of the 
pulsar has brightened over the two years since the previous observation.  By 
2000, the jet is well-defined with structure extending 4$\arcmin$ from the 
pulsar and it has a curved appearance.  The 2003 \emph{Chandra} observation 
shows that part of the jet $\sim1\arcmin$ to the southeast of the pulsar has 
dimmed and the structure at the end of the jet has become less well-defined.  
The persistence of the inner- and outer-arc structures during the entire time 
range, taken with variability of the jet structure and relative intensity, 
indicates that the observed changes are in fact significant and real since it 
is difficult to imagine an artifact that would only apply to discrete parts 
of the images.  Hence, the jet brightening indicates the existence of an 
underlying mechanism, as discussed in \S \ref{sec:disjet}. 

\begin{figure}
\epsscale{1.1}
\plotone{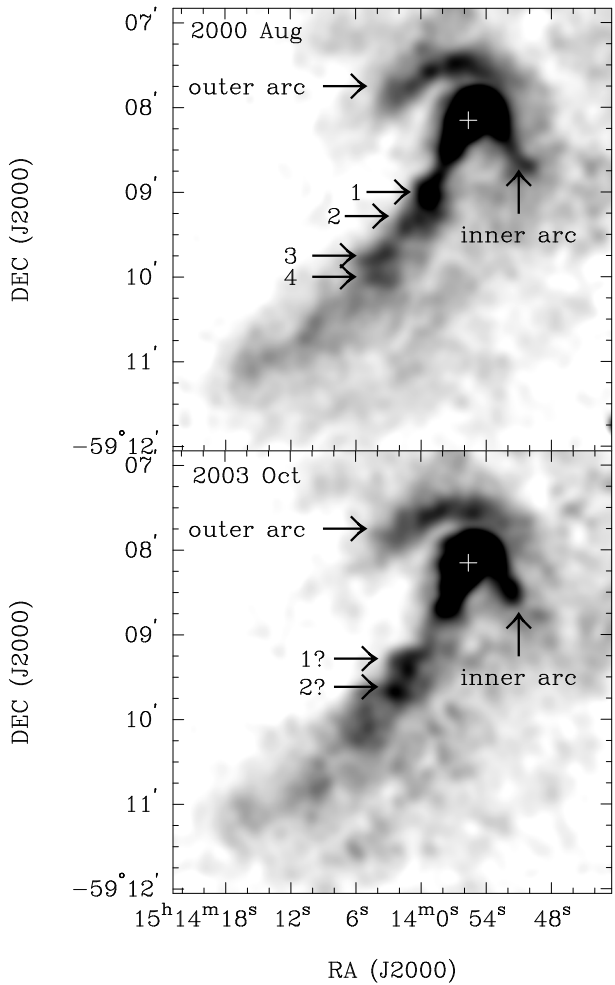}
\caption{\emph{Chandra} images convolved with a Gaussian of FWHM 10$\arcsec$ 
showing changes in the jet.  The numbers indicate the clumps as described in 
the text.  The energy range is 0.3$-$8.0keV.  The grayscale is linear and 
ranges from 0.7\% to 7\% of the peak value.  
\label{chjet}}
\end{figure}

Despite remaining steady in total flux between 2000 and 2003, the jet 
shows variability on $\sim 20\arcsec$ size scales, as shown in Figure 
\ref{chjet}.  In 2000, we identify four large ($20\arcsec$, physical size 
of 0.5 pc) clumps in the jet (labeled as 1$-$4 on the top panel of Figure 
\ref{chjet}) located between $\sim1\arcmin$ (a separation from the pulsar 
of 3 pc at a 30$\degr$ inclination) and $\sim2\farcm5$ away from the pulsar.  
In 2003, there are two $\sim20\arcsec$-sized clumps located $\sim1\farcm5$ and 
1$\farcm7$ away from the pulsar with fainter emission further down the jet.  
If we interpret the clumps as the same structure having moved along the jet, 
then the velocity is $\sim0.5c$ (assuming a 30$\degr$ inclination to the line 
of sight and correcting for relativistic Doppler boosting).  Due to the loss 
of dithering for the 2003~Apr~28 observation, we cannot determine if the jet 
structure changed on week-long time scales.  However, the 10 ks of data in the 
2003~Apr~21 observation shows jet clumps with similar sizes and locations 
as in the 2003~Oct observation indicating that the jet structure changes on 
timescales longer than 6 months.  

\subsection{The Inner and Outer Arcs}
\label{sec:arcs}

\emph{ROSAT} images indicated the presence of a cross-like structure around 
the pulsar \citep{bb97}.  With the improved resolution of \emph{Chandra}, the 
``cross'' was resolved into inner and outer arcs (features ``5'' and ``E'' of 
\citetalias{gak02}) located approximately 17$\arcsec$ and 40$\arcsec$ 
(distances of 0.4pc and 1.2 pc at a 60$\degr$ inclination perpendicular to 
the jet) away from the pulsar.  To determine if the arcs have changed 
structure, we constructed \emph{Chandra} and \emph{ROSAT} HRI images matched 
in energy range, counts, and resolution to the \emph{ROSAT} PSPC images as 
described in \S \ref{sec:xobs} and \S \ref{sec:jet}.  Due to the 
$\sim 25 \arcsec$ resolution of the 
\emph{ROSAT} PSPC detector, emission from the pulsar and small-scale knots 
surrounding the pulsar extends into the inner and outer arc structures.  
Therefore, we could not exclude the pulsar before smoothing without 
adversely affecting the arc structures.  Figure \ref{rcjet} shows the 
comparison between the \emph{ROSAT} and \emph{Chandra} images.  There appears 
to be structural changes to both the inner and outer arcs between 1991/1992, 
1994, 2000, and 2003.  In the same manner as for the diffuse PWN in 
\S \ref{sec:dpwn}, we performed simultaneous absorbed power law spectral 
fits to the outer arc using the \emph{ROSAT} PSPC and \emph{Chandra} data 
holding $N_H$ fixed at $8.6\times10^{21}$~cm$^{-2}$, requiring $\Gamma$ to be 
the same for all the data sets, and allowing the normalization to vary 
between epochs.  For the \emph{Chandra} data, we reproduce the flux and 
photon index reported by \citetalias{gak02} within the errors.  As noted by 
\citetalias{gak02}, the outer arc has the same photon index as the jet and 
the photon index is flatter than for the diffuse PWN.  The outer arc appears 
to have \emph{decreased} in brightness between 1991/1992 and 2000 by 
$\sim$20\% as shown in Table \ref{fluxes}, however this is only a 1$\sigma$ 
result (68\% confidence limit).  This slight flux decrease may just be an 
artifact of the $\sim 25 \arcsec$ resolution of the \emph{ROSAT} PSPC 
detector and the pile-up of the pulsar in the \emph{Chandra} data (the low 
resolution of \emph{ROSAT} results in pulsar counts contaminating the outer 
arc region and the pile-up of the pulsar with \emph{Chandra} results in less 
contamination from the pulsar at the same resolution).

\begin{figure}
\epsscale{1.05}
\plotone{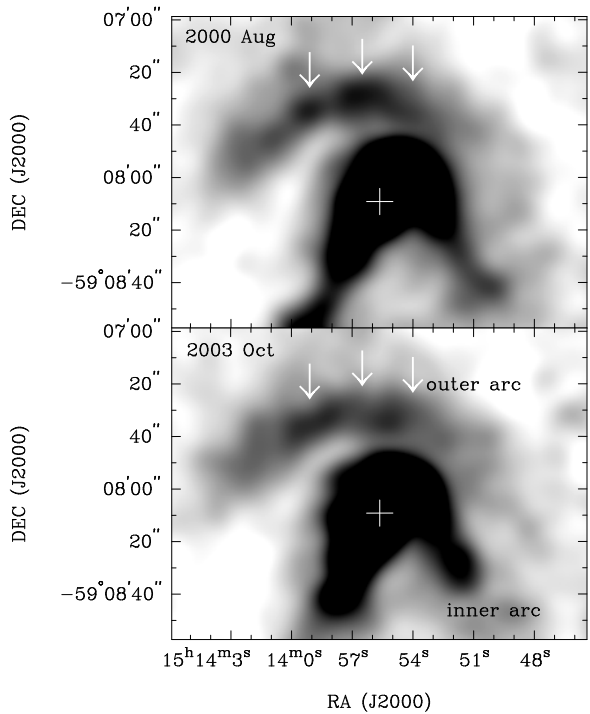}
\caption{\emph{Chandra} images convolved with a Gaussian of FWHM 10$\arcsec$ 
showing changes in the inner and outer arcs over a three-year timescale.  The 
white arrows indicate locations of structural change between 2000 and 2003.  
The energy range is 0.3$-$8.0keV and the exposure time in each image is 
20 ks.  The grayscale is linear and ranges from 1.4\% to 5.9\% of the peak 
value. 
\label{charc}}
\end{figure}

\begin{figure}
\epsscale{1.05}
\plotone{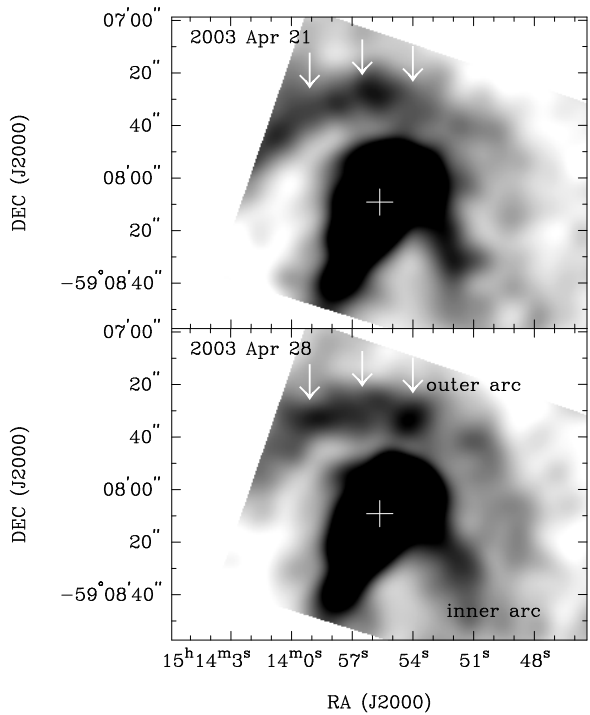}
\caption{\emph{Chandra} images convolved with a Gaussian of FWHM 10$\arcsec$ 
showing changes in the inner and outer arcs over the span of a week.  The 
white arrows are the same as in Figure \ref{charc}.  The energy range is 
0.3$-$8.0keV and the exposure time in each image is 10 ks.  The grayscale is 
linear and ranges from 1.4\% to 5.9\% of the peak value.  The images have 
been blanked to omit chip gap and dead column areas that could not be 
exposure corrected in the 2003~Apr~28 observation.
\label{charc2}}
\end{figure}

If we just consider the \emph{Chandra} data, the outer arc shows time 
variability between 2000~Aug and 2003~Oct as shown in Figure \ref{charc}.  
In 2000~Aug, the outer arc contained two major clumps -- the smaller clump 
having a 10$\arcsec$ diameter, and the larger clump elongated along the arc 
with a size of $10\arcsec\times20\arcsec$ (0.25 pc $\times$ 0.5 pc).  In 
2003~Oct, the clump locations had changed within the outer arc, although the 
clump sizes were comparable to the 2000 clumps.  Similar clump variability is 
also seen in the outer arc between 2003~Apr~21 and 2003~Apr~28 as shown in 
Figure \ref{charc2} indicating that the timescale for arc variability might 
be as short as a few days.  The white arrows on Figures \ref{charc} and 
\ref{charc2} identify locations of transverse structural change between 2000 
and 2003.  Note that the changes observed in the outer arc between the 
2003~Apr images are only slightly greater than one would expect from Poisson 
statistics.  The greatest variation between the two 2003~Apr images, 
indicated by the rightmost arrow on Figure \ref{charc2}, is only 3$\sigma$ on 
unsmoothed images.  Spectral fits indicate that the total brightness of the 
outer arc has remained the same from 2000 to 2003.  

The inner arc also shows a brightening to the west of the pulsar between 
2000 and 2003, as seen in Figures \ref{chjet} and \ref{charc}, however this 
is most likely due to the bright, pointlike source near 
(J2000) R.A. $15^{\mbox{h}}13^{\mbox{m}}51\fs5$,
decl. $-59\degr08\arcmin25\arcsec$ in the 2003~Oct image seen most clearly in 
the bottom panel of Figure \ref{smallscale}.  A SIMBAD search shows no 
variable source in this location.  While it is possible that this source 
could be related to the small-scale knots identified in \S\ref{sec:knots}, we 
believe that it is more likely an artifact of a ``hot pixel'' in the data.  
The southwestern tip of the inner arc does appear to vary slightly between 
2003~Apr~21 and 2003~Apr~28 (see Figure \ref{charc2}), however, this may be 
due in part to the lower signal-to-noise ratio on the 10 ks exposures.

\begin{figure}
\epsscale{1.2}
\plotone{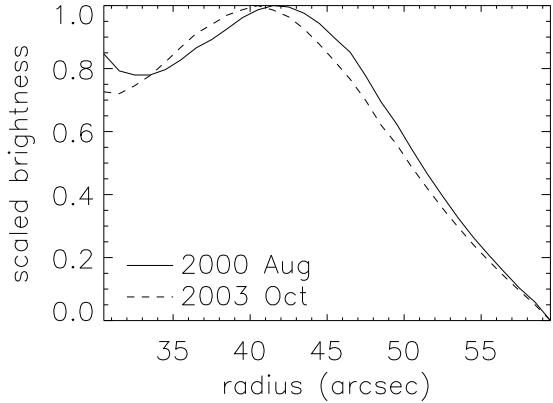}
\caption{Angle-averaged radial profiles for the outer arc for the 2000 Aug and 
2003 Oct \emph{Chandra} images convolved with a Gaussian of FWHM 
10$\arcsec$.  The difference in brightness at a radius of 
$\sim31\arcsec-32\arcsec$ is due to the variation of small-scale, compact 
knots near the inner arc between 2000 and 2003.  Although the small-scale 
knots are unresolved at 10$\arcsec$ resolution, they do contribute 
significantly to the flux of the inner arc region.  Possible \emph{inward} 
motion is indicated.  
\label{arcmov}}
\end{figure}

The inner and outer arcs were predicted to have outward motions of a few 
arcseconds per year \citepalias{gak02}.  To determine the proper motion of 
the outer arc, we convolved the 2000~Aug and 2003~Oct \emph{Chandra} images 
with a Gaussian of FWHM 10$\arcsec$.  We then constructed an angle-averaged 
radial profile over the entire length of the outer arc and centered on the 
pulsar for each epoch, as shown in Figure \ref{arcmov}.  We measured the 
motion of the brightness profiles by minimizing $\chi^2$ of the difference 
between the profiles at each epoch as a function of radial shift and 
amplitude scaling factor.  We used data between radii of 38$\arcsec$ and 
55$\arcsec$ (corresponding to the peak and outside edge of the brightness 
profiles) to measure the motion.  It was necessary to use the peaks of the 
brightness profiles to break the degeneracy between motion and scaling 
differences between epochs.  Data at radii less than 38$\arcsec$ were 
excluded for the motion measurement because of the brightness variations 
caused by the small-scale knots near the inner arc.  Although the small-scale 
knots are unresolved at 10$\arcsec$ resolution, they do contribute 
significantly to the flux of the inner arc region.  The error was determined 
at the 68\% confidence limit of the $\chi^2$ distribution.  We confirmed the 
error by repeating the $\chi^2$ measurement for ten Monte Carlo simulations 
of the Poisson noise and calculating the rms scatter of the minimum $\chi^2$ 
position.  The outer arc appears to have moved \emph{inward} by 
$1\farcs0\pm0\farcs2$ over the 3-year timespan from 2000 to 2003.  This 
corresponds to a velocity of $0.03c$ (assuming outflow perpendicular to the 
jet with a 60$\degr$ inclination to the line of sight and correcting for 
relativistic Doppler boosting).  The convolution of the images and the
angle-averaging of the outer arc results in a measurement of the motion of 
the ensemble average of the material comprising the arc which allows us to 
make a comparison to the model of \citetalias{gak02}.  Thus, we can report 
that the average arc structure appears to move inward.  However, given the 
degree of structural change in this timespan, we cannot definitively 
determine if this represents a real inward motion or is just a result of 
clump variation.  We can rule out unaliased outward motion of the outer arc 
at the 5$\sigma$ level.  For the inner arc, there was too much structural 
change due to the nearby small-scale, compact knots to determine proper 
motions for that feature.

\section{Spatially Resolved Spectroscopy}

Although the jet structure has changed over the 3-year timespan of the 
\emph{Chandra} observations, individual spectral fits to the 2000~Aug and 
2003~Oct data indicate that there was no photon index evolution in the jet 
and diffuse PWN during that time as shown in Table \ref{fluxes}.  To 
determine if the photon index of the diffuse PWN and jet varied spatially, we 
used just the four \emph{Chandra} data sets and performed simultaneous 
absorbed power law fits as outlined in \S \ref{sec:dpwn}.  For each region 
described below, we held $N_H$ fixed to 8.6$\times10^{21}$~cm$^{-2}$, 
required the same $\Gamma$ for all the data sets, but allowed the 
normalization to vary between epochs.  

\begin{figure}
\epsscale{1.2}
\plotone{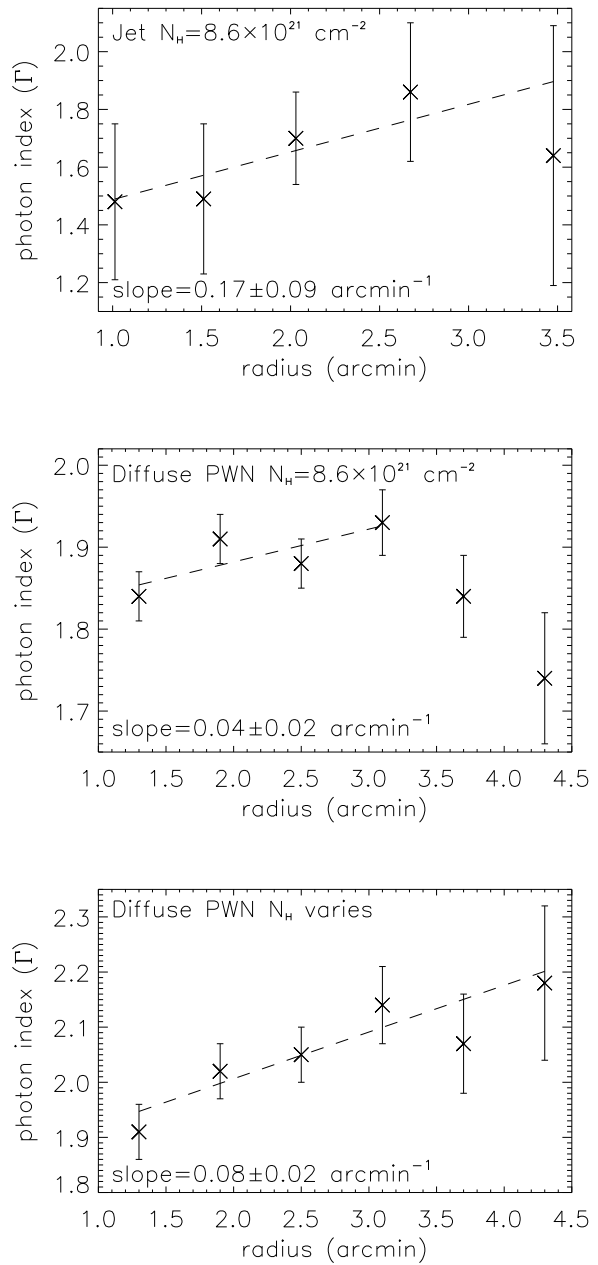}
\caption{Plots of photon index vs. radius for the jet and diffuse PWN with 
linear least-squares fits (dashed lines) and the slope of the fit indicated.  
Error bars are shown for the 90\% confidence limits.  \label{radfig}}
\end{figure}

For the diffuse PWN, we extracted spectra from concentric 36$\arcsec$-wide 
annuli extending from a radius of 1$\farcm$3 to 4$\farcm$3, centered on the 
pulsar as outlined in Table~\ref{radtable}.  The background region used for 
the diffuse PWN was a 36$\arcsec$ wide annulus just exterior to the 
outermost ``source'' annulus and interior to the RCW~89 region.  The 
background annulus was exterior to most of the PWN except to the northwest 
and to the southeast where it includes faint PWN emission.  The jet was 
excluded from the diffuse PWN analysis.  We performed a linear least-squares 
fit to determine the rate at which the photon index changed with radius.  The 
results are shown graphically in Figure \ref{radfig}.  The diffuse PWN shows 
a steepening of 0.04$\pm$0.02~arcmin$^{-1}$ between $1\farcm3$ and 
$3\farcm1$ (for a total $\Delta\Gamma$ of $\sim 0.07$) but flattens 
significantly thereafter.  If we allow $N_H$ to also vary for the diffuse PWN 
fits, the photon index does not flatten at large radii and the best fit slope 
is 0.08$\pm$0.02~arcmin$^{-1}$ which translates to a total change in photon 
index between $1\farcm3$ and $4\farcm4$ of 0.25.  $N_H$ increases from 
9.2$\times10^{21}$~cm$^{-2}$ in the annulus closest to the pulsar to 
13.2$\times10^{21}$~cm$^{-2}$ in the annulus furthest from the pulsar.  These 
values for the absorbing column are consistent with those derived by 
\citet{tmc96} showing that there was about 1.5 times more absorption towards 
RCW~89 than towards the pulsar.

\begin{deluxetable}{lccc}
\tabletypesize{\footnotesize}
\tablecaption{Radial variations in photon index for the diffuse PWN and 
jet.    
\label{radtable}}
\tablewidth{0pt}
\tablehead{
\colhead{} &
\colhead{Distance} &
\colhead{$N_H$} & 
\colhead{} \\
\colhead{Region} &
\colhead{($\arcmin$)} &
\colhead{($10^{21}$ cm$^{-2}$)} &
\colhead{$\Gamma$}}
\startdata
Jet\tablenotemark{a} & 1.0 & 8.6 (fixed) & 1.48$\pm$0.27 \\
    & 1.5 & & 1.49$\pm$0.26 \\
    & 2.0 & & 1.70$\pm$0.16 \\
    & 2.7 & & 1.86$\pm$0.24 \\
    & 3.5 & & 1.64$\pm$0.45 \\
Diffuse PWN\tablenotemark{b} & 1.3 & 8.6 (fixed) & 1.84$\pm$0.03 \\
    & 1.9 & & 1.91$\pm$0.03 \\
    & 2.5 & & 1.88$\pm$0.03 \\
    & 3.1 & & 1.93$\pm$0.02 \\
    & 3.7 & & 1.84$\pm$0.03 \\
    & 4.3 & & 1.74$\pm$0.08 \\
Diffuse PWN & 1.3 & \phn9.2$\pm$0.5 & 1.91$\pm$0.05 \\
    & 1.9 & \phn9.7$\pm$0.5 & 2.02$\pm$0.05 \\
    & 2.5 & 10.3$\pm$0.5 & 2.05$\pm$0.05 \\
    & 3.1 & 10.6$\pm$0.6 & 2.14$\pm$0.07 \\
    & 3.7 & 10.9$\pm$0.8 & 2.07$\pm$0.09 \\
    & 4.3 & 13.2$\pm$1.4 & 2.18$\pm$0.14 \\
\enddata
\tablenotetext{a}{The regions used to extract spectra for the jet are shown 
in Figure \ref{rcjet}.  Distances from the pulsar are measured at the 
geometric center of each region.}
\tablenotetext{b}{The regions used to extract spectra for the diffuse PWN are 
concentric 36$\arcsec$ wide annuli.  Radius is given at the center of each 
annulus.  The jet region has been excluded from the diffuse PWN analysis.}
\end{deluxetable}

For the jet, we constructed five regions along the length of the jet as shown 
in Figure \ref{allim}.  The separation from the pulsar was defined at the 
center of each region.  The background for the jet analysis was chosen as the 
region immediately exterior to the total jet region.  A linear least-squares 
fit was applied to the data and is shown in Figure \ref{radfig}.  The jet 
shows a marginal photon index steepening of 0.17$\pm$0.09~arcmin$^{-1}$ 
between radii of $1\arcmin$ to $3\farcm5$ for a total change of 0.43.

\section{Discussion}
\label{sec:dis}

\subsection{Variability of the Arcs}
\label{sec:disarcs}

\citetalias{gak02} suggested that the inner and outer arcs represent 
compressions downstream of the termination shock in an equatorial 
electron-positron pair outflow.  The compressions are induced as the high 
energy heavy ions (``protons'') embedded in the flow enter the shock heated 
downstream pairs with the same velocity and Lorentz factor as the upstream 
pair wind.  This was an application of the model advanced by \citet{ga94} for 
the time variable wisps seen in the termination shock region of the Crab 
Nebula, which has been shown by \citet[hereafter SA04]{sa04} to give a good 
representation of the structure and toroidally averaged variations seen in 
the inner X-ray ring as well as the propagating wave structures emerging 
from that ring seen by \emph{Chandra} \citep{hmb02}.  In the application of 
that model to G320.4$-$1.2, the inner and outer arcs were interpreted as the 
turning points in the ions' orbits as they do their first gyrations in the 
shock compressed magnetic field.  Fitting the model to the \citetalias{gak02} 
observations of G320.4$-$1.2 led to an estimate of the pair flux, upstream 
flow Lorentz factor, ion flux and magnetic field in the equatorial flow.  
From that, \citetalias{gak02} derived an ion cyclotron time on the order of 
one to two years, which is the basic variability time of the ion induced 
compressions.  If all of the variation is in the form of bulk motion, the 
predicted outflow velocity from \citetalias{gak02} is $\sim 0.5c$.  Although 
our observations do not show an outward velocity, we do confirm the 
theoretical prediction of the variability time scale expected in the inner 
and outer arcs.  That model, being a toroidal average, did not predict the 
angular structure of the arcs uncovered by the new data presented here.  

This model depends on the existence of ultra high energy heavy ions being 
present in the equatorial flow, with the ions carrying a large enough 
fraction of the energy flux to be able to induce substantial compression in 
the radiating pair plasma. The inferred ion number flux, both in the Crab and 
in G320.4$-$1.2, is approximately what one expects on electrodynamic grounds, 
if the ion stream is the electric return current required to prevent the 
charging up of the neutron stars.  It is not currently understood how the 
particles in the flow achieve their high energies ($\gamma\sim10^6$), although 
it has been suggested that they are accelerated and heated by the mysterious 
dissipation processes that lead to the equatorial wind having magnetization 
as weak as has been inferred \citepalias[and references therein]{sa04}.

These kinetic models focus on the time dependent morphology in and around the 
termination shock in the pairs -- in effect, the ions form part of a resolved 
shock structure whose basic form is an equatorial ring. The large scale 
structure of the nebular flow has been addressed in a series of MHD models 
and simulations, whose focus is on understanding the dramatic pole to 
equator asymmetry, especially the appearance of the jets 
\citep{bk02,lyu02,kl03,kl04,dab04,bck05}. These models assume all the 
particles have small Larmor radius, and thus are indifferent to composition 
and exclude the proposed high energy ions by assumption. With the premise 
that the injected energy flux decreases with increasing rotational latitude 
$\theta$ (usually assumed proportional to $\cos^2 \theta $), the energy is 
mostly injected through a toroidal belt shock, with some flow passing 
obliquely through a higher latitude ``arch shock'' which, because the shock 
is oblique to the flow, does not decelerate the plasma to subsonic speeds.  
These non-spherically symmetric shock structures can create the appearance of 
an inner ring of emission, with the weakly decelerated plasma emerging from 
the arch shocks contributing to form the appearance of an outer ring.  Doppler 
boosting may yield an appearance of partial arcs rather than complete 
rings\footnote{However, such a model then has difficulty explaining why the 
inner X-ray ring in the Crab Nebula has an approximately axisymetric 
appearance, after toroidally averaging over the knots in the ring brightness, 
while the larger radius torus does show signs of Doppler boost.  The ion 
compression model has the same difficulty. One of the major successes of the 
MHD model is to find a flow that does have substantial Doppler boost at 
distances from the pulsar comparable to the Crab torus}.  These models 
provide a plausible scenario for the formation of the jets, as the 
consequence of backflow and magnetic hoop stress at higher latitudes.

The simulations of this model show no signs of short time variability
corresponding to the wisps\footnote{\citet{bck05} do report a long period 
quasi-coherent oscillation of the flow ($t \sim 40$ years) with some possible 
shorter time variability when the magnetization is finite but not large -- the 
variability vanishes for unmagnetized flow models.  The physical reasons for 
this variability are not understood (Bogovalov, private communication).}.  
The simulations may lack sufficient resolution to find the Kelvin-Helmholtz 
instability between the outward flow emerging from the arch shock and the 
higher latitude backflow, an effect suggested by \citet{beg99} with a somewhat 
different flow geometry in mind as the origin of the wisps.  Alternatively, 
magnetization high enough to form the jet may suppress the shear flow 
instabilities.

While the effects of high energy equatorial ions have not been explicitly 
included in the MHD simulations, if present they would be injected into the 
shock decelerated pairs through the equatorial ring shock seen in the MHD 
simulations.  Thus the formation of ion driven compressions is qualitatively 
consistent with the MHD models, and remains the only quantitatively 
elaborated model for the time variability seen in the Crab wisps and perhaps 
seen in the data described here. Based on this model, \citetalias{gak02} 
predicted arc motions of a few arcseconds per year.  We see no outward motion 
of the outer arc and perhaps even an inward motion.  \citetalias{sa04} noted 
that the positions of the ion driven compressions can move inward as well as 
outwards, with the apparent motion then depending on when one takes a 
snapshot of the structure -- the possible existence of apparent inward motion, 
perhaps seen in our data, is a prediction of the model.  However, the 
apparent motion may also be due to the transverse structural variations in 
the outer arc between 2000 and 2003 -- non-axisymmetric variability, a 3D 
effect, has not been included in any of the models published to date. If such 
transverse change is the origin of the apparent motion, the arcs would then 
be quasi-stationary in the radial sense.  Or, we could be witnessing alaising 
as seen by \citet{sca69} in his observations of the wisps in the Crab, due to 
the observations having undersampled the variations.

We can test whether the quasi-stationary nature of the outer arc is due to 
aliasing.  If we assume that the particles comprising the outer arc are 
relativistic, and thus Doppler boosted, then we can determine a speed based 
on the brightness ratio of $\ga5$ between the near and far sides.  For a 
photon index of 1.6, the Doppler boosting formula reduces to 
$\beta \cos\phi=0.22$, where $\beta=v/c$ and $\phi$ is the inclination angle 
to the line-of-sight.  If the inclination of the outer arc is 
60$\degr$ (oriented 90$\degr$ from the jet), then 
$\beta=0.44$.  At this speed, we expect the outer arc to move nearly 
16$\arcsec$ in 3 years, which is slightly greater than the width of the arc.  
Over 6 months, the outer arc should move 2$\arcsec-3\arcsec$, which is not 
observed.  Suppose that the motion of particles \emph{through} the arc is not 
indicative of the motion of the arc.  In this case consider that the arc 
would need to move at 0.27c to move its width (0.25 pc) in 3 years.  At this 
speed, the arc would only move 1$\farcs$6 in 6 months.  This amount of motion 
is difficult to rule in or out given the degree of structural change in the 
outer arc and the signal-to-noise ratio of the data.

The nature of arc a and arc b is unknown.  If indeed the arcs are 
outward-moving structures, then perhaps arc a and arc b are just more evolved 
versions of the inner and outer arcs.  They could also represent a 
quasi-coherent oscillation of the flow in the outer PWN as observed in the 
simulations of \citet{bck05}.  A more thorough analysis of these faint PWN 
structures will be performed at a later time.
 
\begin{deluxetable}{cccc}
\tabletypesize{\footnotesize}
\tablecaption{Alfv\'{e}n Crossing Times in the Jet    
\label{alfven}}
\tablewidth{0pt}
\tablehead{
\colhead{} &
\colhead{$B_{\mathrm{min}}$} &
\colhead{$U_{\mathrm{min}}$} &
\colhead{$t_A$\tablenotemark{b}} \\
\colhead{Position\tablenotemark{a}} &
\colhead{($\mu$G)} &
\colhead{(10$^{-11}$ erg cm$^{-3}$)} &
\colhead{(years)}}
\startdata
1 & 15 & 2.1 & 1.3 \\
2 & 17 & 2.8 & 1.3 \\
3 & 14 & 1.7 & 2.0 \\
4 & 12 & 1.3 & 2.0 \\
5 & 11 & 1.1 & 1.6 \\
\enddata
\tablenotetext{a}{Positions are shown in Figure \ref{allim}, position 1 is 
closest to the pulsar.  Volumes at each position are assumed to be 
cylindrical.}
\tablenotetext{b}{The Alfv\'{e}n velocity at each position along the jet as 
computed from Equation \ref{va} is $\approx0.63c$.}
\end{deluxetable}

\subsection{Variability in the Jet}
\label{sec:disjet}

MHD sausage or kink instabilities are viable candidates for the jet 
variability, as in the Vela jet \citep{ptk03}.  The time scale 
for such phenomena is set by the Alfv\'{e}n crossing time 
$t_A = r_{jet} /v_A$, where $r_{jet}$ is the cylindrical radius of 
the jet and
\begin{equation}
\label{va}
v_A = \frac{c}{\sqrt{1+ \frac{4\pi (\rho c^2 + 4p)}{B^2}}} 
      \approx \frac{c}{\sqrt{1+\frac{16 \pi U}{3B^2}}}.
\end{equation}
Here $p$ is the relativistic pressure, assumed to be isotropic, and 
$U = 3p \gg \rho c^2$ is the energy density of the relativistic plasma.  
Shown in Table \ref{alfven} are the magnetic fields, plasma energy densities, 
and Alfv\'{e}n crossing times for regions along the jet as shown in 
Figure \ref{allim}.  Position 1 is closest to the pulsar and we are assuming 
a cylindrical volume.  The magnetic fields and energy densities are computed 
from equipartition arguments using integrated fluxes for the energy range 
0.5-10 keV and assuming an uncooled spectrum \citep[pp. 170-171]{pac70}.  For 
all regions $v_A\approx0.63c$ and the Alfv\'{e}n crossing times vary from 
1.3$-$2 years.  Thus the MHD instability of a magnetic pinch, the basic 
structure of the jet described in the MHD models, certainly is a viable 
candidate for the jet variations observed between 2000 and 2003, and perhaps 
may be a candidate to explain the partial brightening of the jet from 
1991/1992 to 1994 and 2000.  Indeed, episodic jet outflows are observed in 
MHD simulations for certain magnetic field configurations \citep{ops97, op97}.

If we interpret the jet structural changes between 2000 and 2003 as clump 
motion along the jet, then the resulting velocity is $\sim$ 0.5c.  This 
velocity is consistent with the prediction of \citetalias{gak02} based on 
Doppler boosting and also matches the outflow speeds observed in the Crab 
Nebula jet \citep[0.4c,][]{hmb02}, the Vela PWN jet 
\citep[$0.3c-0.7c$,][]{ptk03}, and in the jet of G11.2$-$0.3 
\citep[$0.8c-1.4c$,][]{rtk03}.  Relativistic jet outflow velocities of 
$\sim0.5c$ are predicted by the MHD simulations described above 
\citep{bck05,dab04,kl03,kl04}.  The jet brightnening from 1991/1992 to 1994 
and 2000 cannot simply be a result of material moving at $0.5c$ down the jet 
causing the jet to lengthen and brighten with the inclusion of more 
material.  This scenario is excluded for two reasons, the first is that it 
would take nearly 80 years for material moving at $0.5c$ to travel the length 
of the jet, and the second is that the jet seems to brighten in a ``patchy'' 
manner and does not simply lengthen, as shown in Figure \ref{rcjet}.  
Therefore, as discussed above, the brightening of the jet is more likely the 
result of an MHD instability such as a magnetic pinch.  In this model, the 
instability causes compression and brightening of local regions, each of 
which has length no more than the jet diameter. There is no necessity for the 
energy to be injected into the jet at a varying rate, with the observed 
variations reflecting transport along the jet.  This scenario works because 
the Alfv\'{e}n transit time {\it across} the jet is much shorter than the 
flow time along the jet. The model suggests that the clumps vary incoherently 
with respect to each other, on a time scale approximately equal to the 
Alfv\'{e}n crossing time ($< 2$ years), consistent with the very limited time 
series available to us.

\subsection{Small-Scale Structure Near the Pulsar}

The compact, small-scale knots near the pulsar are quite variable on 6 month 
timescales.  Although the knot to the southeast of the pulsar might represent 
material moving at 0.6c, given the startling knot variability to the 
northwest, a conclusion of knot motion is perhaps premature.  We can estimate 
the magnetic fields in the knots in three ways:  1. assume that the knot size 
is approximately the Larmor radius of gyrating pairs, 2. use 
equipartition/minimum energy, and 3. use synchrotron lifetime if the knots 
are in motion at 0.6c and cool below detectability at a distance of the inner 
arc in time frames from 6 months to 1 year.  For the first method, 
\citetalias{gak02} derived a magnetic field of 3$\mu$G from the Larmor radius 
for their knot~1.  For the second method, the magnetic field derived from 
minimum energy using a size of 0.1 pc and the integrated flux of knot~1 from 
\citetalias{gak02} for the energy range of 0.5$-$10.0~keV is 
$B_{\mathrm{min}}=76\mu$G.  For the third method, the synchrotron lifetime 
can be calculated using (from equation 4 of \citetalias{gak02})
\begin{eqnarray}
t_{\mathrm{synch}}=39B^{-3/2}\varepsilon^{-1/2}~\mathrm{kyr}
\end{eqnarray}
where $B$ is the magnetic field in $\mu$G and $\varepsilon$ is the energy 
in keV.  For 5 keV and $t_{\mathrm{synch}}=1$ year, $B=700\mu$G.  Given this 
high magnetic field, it seems unlikely that synchrotron cooling could be 
responsible for the transient nature of the knots.

The knots may have a similar explanation as ``knot 2'' in the Crab Nebula 
which has been interpreted as an unstable quasi-stationary shock in the polar 
jet outflow \citep{hmb02}.  In this scenario, one would expect the bulk of 
the knots to be in the approaching jet since the receding jet is Doppler 
boosted down in brightness.  However, in this case we observe most of the 
knot structure on the receding jet side.

Given that the knots are never observed beyond the inner arc radius and most 
of them occur on the approaching side of the arcs, it is likely that the 
knots are either associated with the equatorial outflow or the resulting 
backflow as described in \S \ref{sec:disarcs} and shown in Figure~3 of 
\citet{kl03}.  Two processes expected from the relativistic MHD simulations 
that could produce ``knot-like'' emission are Kelvin-Helmholtz (KH) 
instabilities arising in the shear layer between the equatorial outflow and 
the backflow and unsteady vortices that develop in the converging flow at the 
base of the jet.  

The KH instability creates waves that may grow and break, while traveling in 
the direction of the mass weighted jump in velocity across the interface 
between the outflow and backflow.  Doppler boosting as the breaking waves 
head toward us, and plasma and B field compression as the waves break, may 
lead to the formation of bright features that can appear to us as localized 
bright spots of emission.  The main difficulty with this idea is that 
within the flow models, the mass weighted velocity jump is directed toward us 
only on the near (visible jet) side except very close in to the base
of the jets.  This is because flow convergence increases the mass density in 
the layer that converges on the axis at small distance from the axis, while 
the outflowing layer has low density at small radii. At larger radii, the 
relative densities of the layers reverses.  Thus this effect will work 
only if there is a lot of flow convergence.

The converging flow at the base of the jet could also produce transient 
features due to the development of unstable vortices (N. Bucciantini, private 
communication).  The backflow would need to focus quite well toward the base 
of the jet to produce Doppler boosting of flows directed towards us and may 
be rather sensitive to the maintenance of axisymmetry.  However, this 
hypothesis naturally explains the relative absence of knots on the near 
(visible) jet side since the converging flow at the base of the approaching 
jet would be directed away from us.

Regardless of the actual cause of the knots, they seem to be 
diagnostic of flow details in the PWN.  To determine the true nature 
of the knots will require much higher resolution simulations and more 
analysis of the flow models.

\subsection{Synchrotron Cooling in the Jet and Diffuse PWN}

\citetalias{gak02} determined that the synchrotron cooling time for the jet 
at 5 keV and a magnetic field of $\sim25~\mu$G is 140 years.  This cooling 
should result in a $\Delta\Gamma$ of 0.5 \citep{kar62,ps73}.  Indeed, there 
is a marginal indication of cooling in the jet with a total $\Delta\Gamma$ of 
0.43.  We await analysis of the new, longer \emph{Chandra} observation to 
verify the possible synchrotron cooling in the jet.  The relativistic 
MHD simulations do predict that the jet should undergo synchrotron cooling, 
although there is mixing with the ambient PWN material (S. Komissarov, 
private communication).  At the base of the jet, the photon index should 
match that of the equatorial 
outflow since the jet is constructed from purloined equatorial material.  If 
we assume that the inner and outer arcs identify the equatorial outflow 
material, then their photon index of 1.6 does nearly match the photon index 
of 1.5 at the base of the jet.  If the spectral steepening along the jet is 
real, it would be similar to that observed in the Crab Nebula jet 
\citep[$\Delta\Gamma\sim0.5$;][]{mbh04}.

Spectral steepening with increasing radius has been observed in a number of 
PWNe and is generally attributed to both synchrotron cooling and expansion 
losses \citep{gs03,lwa02,mbh04,shp01,shs04}.  The degree of steepening 
varies from $\Delta\Gamma\sim0.3$ to 1.3 with a typical value of 1 and the 
radial photon index profiles generally show a monotonic increase with radius.  
Our spectral results for the diffuse component of the PWN of G320.4$-$1.2 
also show a monotonic increase of photon index with radius if we allow the 
absorption to vary, but the degree of steepening ($\sim0.25$) is less than in 
other PWNe.  In contrast, models for synchrotron cooling in PWNe based on 
the \citet{kc84} model predict a 
rapid increase in photon index at large radius \citep{r03,shs04}.  Perhaps the 
discrepancy between the cooling model and the observations is due to 
large-scale mixing of recently accelerated material with ``older'' material 
as predicted by the relativistic MHD models 
\citep{bk02,lyu02,kl03,kl04,dab04,bck05}.

\section{Summary and Conclusions}
\label{sec:con}

Variability is observed in the X-ray PWN of PSR B1509$-$58 on time scales 
possibly as short as one week and up to twelve years.  Our primary results 
are as follows:

\begin{enumerate}
\item{The compact, small-scale knots appearing within 20$\arcsec$ of the 
pulsar exhibit transient behavior which may be attributed to turbulence in 
the flows surrounding the pulsar.  Possible knot motion is 
indicated with a velocity of $0.6c$.} 
\item{Apparent outflow along the jet is observed with velocities of 
$\sim0.5c$.  This outflow alone cannot account for the $\sim$30\% 
brightening of the jet between 1991/1992 and 2000.  The Alfv\'{e}n crossing 
time for the jet is 1.3$-$2 years, therefore, MHD kink or sausage 
instabilities can account for the rapid morphological variations and perhaps 
the partial jet brightening.}
\item{The outer arc has possibly moved inward with a velocity of $0.03c$, 
however the transverse structural changes seen in the outer arc may account 
for the apparent motion.  We cannot determine at this time if the outer arc 
is truly quasi-stationary or if we are witnessing aliasing.}
\item{The diffuse PWN has not evolved significantly in structure or 
brightness over the 12-year time span.  Using the summed \emph{Chandra} 
images, we identify two possible arc structures exterior to the outer arc.}  
\item{The photon indices of the diffuse PWN and possibly the jet steepen with 
increasing radius indicating synchrotron cooling at X-ray energies.}
\end{enumerate}

Although our imaging capabilities have improved substantially since the first 
optical observations of time variability in the Crab Nebula \citep{sca69}, 
our understanding of these variations in PWNe is still quite limited.  For 
instance, while we expect magnetic fields to play an important dynamical role 
in jets, and indeed we do see variations on the appropriate Alfv\'{e}n 
crossing times, we do not know for certain if MHD instabilities are the root 
cause of the observed variations.  The arc structures we observe in 
G320.4$-$1.2 are equally enigmatic.  We do not yet know if they are in steady 
motion or are quasi-stationary wave phenomena.  The striking changes in the 
small-scale knots near the pulsar may simply be ``weather,'' diagnosing 
unimportant details in the PWN flow, or they may indicate important flow 
structure which is essential to understanding, for instance, diffusion of 
particles from the equatorial flow to higher latitudes, a loss essential to
a post pair shock second-order Fermi acceleration model.  Certainly deeper, 
and appropriately spaced, X-ray observations will help resolve some issues 
such as the possible spatial aliasing of the outer arc.  Also, the higher 
signal-to-noise will provide better constraints on the spatial spectral 
index variations and allow us to determine if and how much mixing has 
occurred in the diffuse PWN.  Finally, we are excited by the recent 
development of relativistic MHD models and we hope that some of the 
variability we observe here can eventually be observed in those simulations. 

\acknowledgments

We thank Fred Seward for assistance with \emph{ROSAT} data analysis, Elena 
Amato, Niccol\`{o} Bucciantini, and Serguei Komissarov for helpful 
discussions, and the anonymous referee for a careful reading of the 
manuscript.  This research has made use of data obtained from the High 
Energy Astrophysics Science Archive Research Center (HEASARC), provided by 
NASA's Goddard Space Flight Center.  T. D. \& B. M. G. acknowledge the 
support of NASA through SAO grant GO3-4063A.  J.A. acknowledges support from 
NASA Chandra theory grant TM4-5005X, NASA ATP grant NAG5-12031, and from the 
taxpayers of California.

\end{document}